# BEAMED CORE ANTIMATTER PROPULSION: ENGINE DESIGN AND OPTIMIZATION

Ronan L. Keane

*Western Reserve Academy, 115 College Street, Hudson, Ohio 44236, USA*

**keaner@wra.net**

Wei-Ming Zhang

*Department of Physics, Kent State University, Kent, Ohio 44242, USA*

**wzhang@kent.edu**

**SUMMARY**

A conceptual design for beamed core antimatter propulsion is reported, where electrically charged annihilation products directly generate thrust after being deflected and collimated by a magnetic nozzle. Simulations were carried out using the Geant4 (Geometry and tracking) software toolkit released by the CERN accelerator laboratory for Monte Carlo simulation of the interaction of particles with matter and fields. Geant permits a more sophisticated and comprehensive design and optimization of antimatter engines than the software environment for simulations reported by prior researchers. The main finding is that effective exhaust speeds $v_e \sim 0.69c$ (where $c$ is the speed of light) are feasible for charged pions in beamed core propulsion, a major improvement over the $v_e \sim 0.33c$ estimate based on prior simulations. The improvement resulted from optimization of the geometry and the field configuration of the magnetic nozzle. Moreover, this improved performance is realized using a magnetic field on the order of 10 T at the location of its highest magnitude. Such a field could be produced with today's technology, whereas prior nozzle designs anticipated and required major advances in this area. The paper also briefly reviews prospects for production of the fuel needed for a beamed core engine.

1.  **INTRODUCTION**

While antimatter has been a rich source of inspiration for writers of science fiction, it has also garnered considerable attention in the astronautical engineering literature, where many conceptual studies have considered antimatter as a fuel for spacecraft propulsion [1-35]. The task of producing and safely storing antimatter in macroscopic quantities may prove to be unfeasible or prohibitively difficult, and will undoubtedly be a very futuristic technology if it ever comes to fruition. On the other hand, the incomparable energy storage per unit mass of this potential fuel motivates a very long-term perspective when considering feasibility. The nominal energy released per kilogram of annihilating antimatter and matter is $9 \times 10^{16}$ joules — about two billion times larger than the thermal energy from burning a kilogram of hydrocarbon, or over a thousand times larger than liberated from a kilogram of fuel in a nuclear fission reactor.

It is well known that in order to maximize the final speed of a rocket, it is necessary to consider only exhaust speed ($v_e$), the fraction of the initial mass devoted to fuel, and the configuration of stages. The latter two factors depend strongly on fine details of engineering and construction, and when considering space propulsion for the distant future, it seems appropriate to defer the study of such specifics. Thus, exhaust speed is seen as the main focus of long-term advances in propulsion technology.

The particular category of antimatter propulsion studied in this paper, called the "beamed core" concept, uses the relativistic charged particles (mostly pions) produced in antiproton-proton annihilation to directly generate thrust, after deflection by an electromagnetic nozzle. Because of the high relativistic velocities of the charged mesons directly produced in these annihilation reactions, the beamed core concept offers very high $v_e$ performance, but is still subject to inefficiencies in converting the stored antimatter energy into propulsive momentum transfer. For example, much of the energy goes into electrically neutral particles which contribute nothing to the thrust, and the nozzle has a limited efficiency for deflecting and collimating the charged particle exhaust.

There are antimatter-based alternatives to the beamed core engine design. The thermal antimatter rocket concept harnesses a larger fraction of the stored antimatter energy than in the beamed core case, and allows a non-exotic material to comprise a very large fraction of the propellant, but $v_e$ is much lower. The approach of using antiprotons to catalyze a fission/fusion reaction has many variations [15-23], and in terms of $v_e$ and required quantities of antimatter, these variations fall somewhere between the thermal antimatter rocket and the beamed core concept. The alternatives to the beamed core engine design offer the possibility of feasible vehicles with, in some cases, vastly smaller amounts of antimatter (micrograms or less).

## 2. PRIOR WORK AND THE STARTING POINT FOR THIS STUDY

The present study represents the second phase of an updated and fully quantitative simulation of beamed core propulsion using the Geant4 software toolkit [36] developed and maintained by the CERN accelerator laboratory for Monte Carlo simulation of the interaction of particles with matter and fields. The primary application of Geant is for design and simulation of particle detectors, and this project is the first known application of Geant to beamed core propulsion. The first phase addressed questions related to substituting a range of matter nuclei instead of the usual hydrogen target in the antimatter on matter annihilation reaction [35], and also used more modern data to infer that the effective exhaust speed for charged pions in a beamed core engine would be $v_e = \langle v_{\pi\pm} \rangle f_n = 0.81c\, f_n$ [35], where $\langle v_{\pi\pm} \rangle$ is the mean speed of charged pions upon emission from the annihilation point, and $f_n$ is the nozzle efficiency, which is most conveniently defined by the equation above. Note that hereafter, the notation $v_e$ refers to charged pions, which make up a large fraction of the collimated exhaust particles [35]. A simulation to determine attainable numerical values of $f_n$ was beyond the scope of Ref. [35] and was deferred to the present second phase of the simulation study. The estimate $\langle v_{\pi\pm} \rangle = 0.81c$ [35] represents a significantly lower performance estimate than the value $\langle v_{\pi\pm} \rangle = 0.92c$, first reported by Morgan [1-3] in 1974, and used consistently in the literature since then, notably by Frisbee [29-33].

The most detailed studies of beamed core propulsion are those by Frisbee [29-33]. These papers rely on Morgan's $\langle v_{\pi\pm} \rangle$ estimates as noted above, and also assume a nozzle efficiency $f_n \sim 0.36$ based on Monte Carlo simulations carried out by Callas in the late 1980s [12]. The magnetic nozzle is the main propulsion component of a beamed core antimatter engine, and optimization of its performance is crucial. The main purpose of the present second phase of Geant-based simulations is to revisit the topic of nozzle efficiency using a modern simulation environment, which permits a much more detailed investigation of nozzle design compared with what was feasible in the 1980s; in particular, Geant4 facilitates a comprehensive scan of the full parameter space associated with the propulsion performance of the family of nozzles under consideration.

Finally, the present paper includes a discussion of the connection between $v_e$ and the possible speed of a future vehicle powered by a beamed core engine. Based on current knowledge, the prospects for production of the needed fuel are also briefly reviewed.

## 3.  THE CHIPS MODEL AND GEANT

The Chiral Invariant Phase Space (CHIPS) event generator [37, 38] models the interactions of various particle types, including antiprotons, with nuclei ranging from hydrogen to uranium. In CHIPS, when an antiproton annihilates on the nuclear periphery, some secondary mesons are absorbed by the nucleus, producing an intranuclear hadronic excitation (quasmon) – a compound of the meson and a cluster of nucleons. Subsequently, the quasmon dissipates energy by quark fusion or quark exchange. The CHIPS event generator has tunable parameters and has been tested against the best available measurements of antiproton annihilations at rest on nuclei. The focus of the CHIPS authors is on emission of relativistic particles: pions, kaons, and light nuclear fragments. There is a good level of agreement in the spectra for emitted relativistic particles like pions, kaons, protons, neutrons and light nuclei up to $^4$He [37-46], and deviations from the experimental measurements are typically on the order of 10%.

The CHIPS event generator is implemented in the CERN Geant4 tool kit [36]. Geant allows realistic three-dimensional magnetic field maps to be defined, and it handles the computation of charged particle trajectories and secondary interactions throughout the spatial region of interest. These features are obviously very helpful for study and optimization of magnetic nozzle design. The use of the CHIPS event generator in the Geant4 software framework offers an ideal environment for more detailed exploration of beamed core antimatter propulsion than previously possible.

## 4. ASSUMPTIONS CONCERNING NOZZLE DESIGN

The magnetic nozzle is designed to deflect electrically-charged reaction products as close as possible to a direction anti-parallel to the thrust vector. If a Cartesian coordinate system is defined with the minus $z$ axis pointing along the thrust direction, and if a charged particle is *initially* emitted from the annihilation point with momentum components ($p_x$, $p_y$, $p_z$), then one of the functions of the nozzle is to reverse the sign of $p_z$ if it is negative. In addition, the field should collimate the charged particles, i.e., minimize the magnitude of the $x$ and $y$ components of the momentum at the exit region of the nozzle. Because no electromagnetic field configuration can perform these function perfectly, the momentum imparted to the vehicle will only be a certain fraction $f_n$ of the sum of the momentum magnitudes of the charged reaction products. The fraction $f_n$ is called the nozzle efficiency. Given that the great majority of the charged particles are pions, this definition is equivalent to the one stated earlier, namely $f_n = v_e / \langle v_{\pi\pm} \rangle$.

For this study of magnetic nozzle design, the first assumption is that a modified solenoid with a varying number of turns per length, $n$ (and/or varying current, $I$) can generate a range of possible magnetic field configurations with the desired property of being strongest at the forward end of the nozzle, becoming progressively weaker towards the aft end of the nozzle. A normal solenoid produces a uniform magnetic field $B = \mu_0 n I$ pointing along the axis of the coil (assumed to be along the $z$ axis), where $\mu_0 = 4\pi \times 10^{-7}$ Tm/A. A parameter $g$ is introduced such that $\mu_0 n I = B_{max}(1 - gz)$. Thus $g$ controls how steeply the field strength varies along the axis of the coil. This type of field has the desired property of being able to reverse the direction of charged particles. The non-uniform solenoid has both radial and axial magnetic field components. The Biot-Savart equation shows that these have the form $B_r(r) = B_{max} \, r \, g/2$ and $B_z(z) = B_{max}(1 - gz)$.

Within the model of this non-uniform solenoid, the nozzle efficiency depends on five parameters: the length and radius of the coil, $L_{coil}$ and $R_{coil}$, the distance of the annihilation point from the forward end of the nozzle, $L_{annih}$ (expressed as a percentage of $L_{coil}$), and also the two magnetic field parameters introduced above: $B_{max}$ and $g$. Since $g$ is an inconvenient parameter with the dimensions of (length)$^{-1}$, instead this study uses $B_{min}$, the magnetic field at the aft end of the nozzle.

There is also a sixth parameter that is relevant – the kinetic energy of the incoming antiproton, $T_{in}$. This study does not explore $T_{in} > 10$ MeV. While even small hospital accelerators easily go above 10 MeV, an antimatter engine needs to inject massive quantities of antimatter, and exceeding this limit might be an engineering challenge. Another assumption is that the fringe magnetic field beyond the interior volume of the coil can be neglected. In the present Geant simulation, this field is set to zero. Geant has the capability to accurately simulate secondary interactions in the material of the coil and in all matter within the chosen volume, but it is assumed here that such effects are negligible.

To limit the number of parameters under study, nozzle efficiency calculations are limited to antiproton on proton annihilations – unlike in Ref. [35], antiprotons on heavier matter nuclei are not considered.

## 5. NOZZLE OPTIMIZATION RESULTS

The 6-dimensional parameter space explained in the previous section, namely $L_{coil}$, $R_{coil}$, $L_{annih}$, $B_{max}$, $B_{min}$ and $T_{in}$ in principle can be very complex, and difficult to map out. This study adopts the following approach to simplify the problem. It is assumed that there is only one peak region in nozzle efficiency, as opposed to multiple peaks, and it is assumed that in the general vicinity of that single maximum, it is possible to locate the optimum settings by scanning the parameters one at a time. In the course of many explorations of the parameter space, no evidence of multiple peaks was found. The CPU time per annihilation event is such that statistical errors are a concern. Therefore, when needed, additional statistics were accumulated near the maximum.

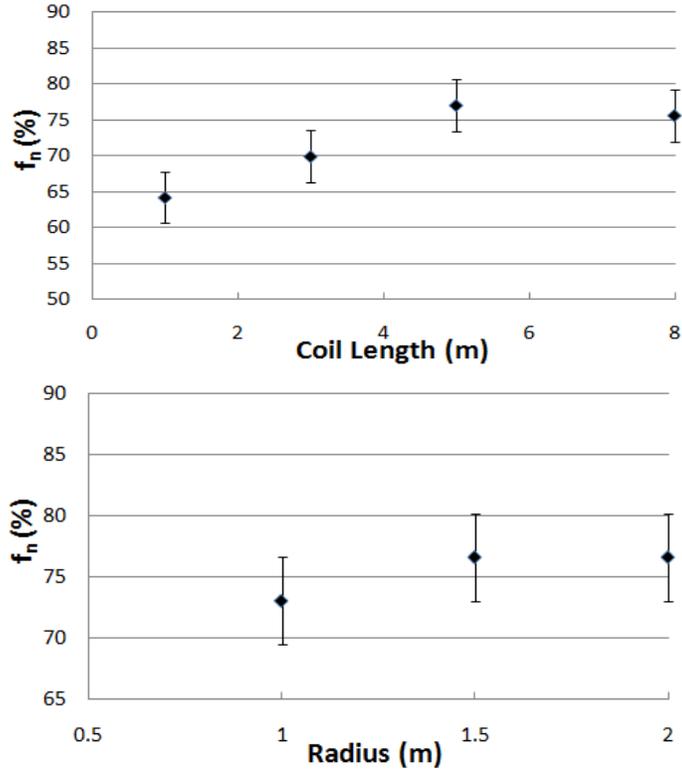

Fig 1: Nozzle efficiency as a function of the length (upper panel) and radius (lower panel) of the modified solenoid. With these two parameters, it is expected to see a plateau, not a peak, because once the charged particle trajectories are fully contained, there is no need to further increase the volume of the nozzle. Statistical errors are large in these two plots, because Geant's event display (see later) already gives a clear picture of when the nozzle is big enough. For these two plots, $B_{max}$ = 10 T, $B_{min}$ = 0 T, and $L_{annih}$ = 75%. For all subsequent tests, $L_{coil}$ = 3.8 m and $R_{coil}$ = 1.5 m.

The incoming antiproton kinetic energy $T_{in}$ is found to only weakly affect the nozzle efficiency over the allowed range (up to 10 MeV). It was set at 10 MeV throughout the exploration that follows. The optimum direction of the incoming antiproton was to the aft.

The conclusion from Fig. 1 above is that a nozzle ~4 m in length and ~1.5 m in diameter is sufficient. The earliest conceptual sketch of a beamed core nozzle found in the literature is by Forward [7]. His nozzle has three large current loops of increasing but unspecified diameter, centered on a common axis, with the aft two separated by 21 m and with the fringe field extending over 1000 m beyond the last loop. In contrast, Callas, who has published the only fully quantitative prior design [12], assumed a single current loop of radius 0.1 m, but with sufficient current to produce a maximum field of 138 T, which is well beyond the reach of today's technology [47]. His nozzle efficiency, as used in calculations by Frisbee [30-33], was $f_n \sim 36\%$.

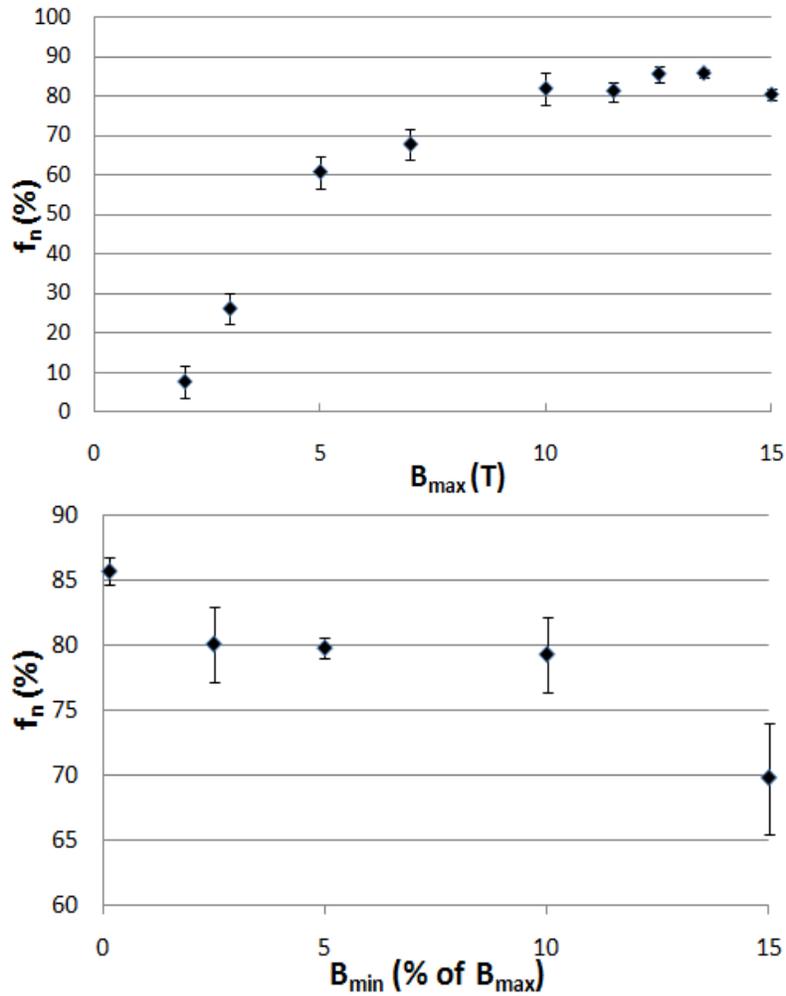

Fig 2: Nozzle efficiency as a function of the maximum (upper panel) and minimum (lower panel) magnetic field. For both plots, $L_{annih} = 75\%$, while $B_{min} = 0$ for the upper plot and $B_{max} = 12.5$ T for the lower plot. The best efficiency is obtained with a steeply varying field that drops close to zero at the aft end of the nozzle.

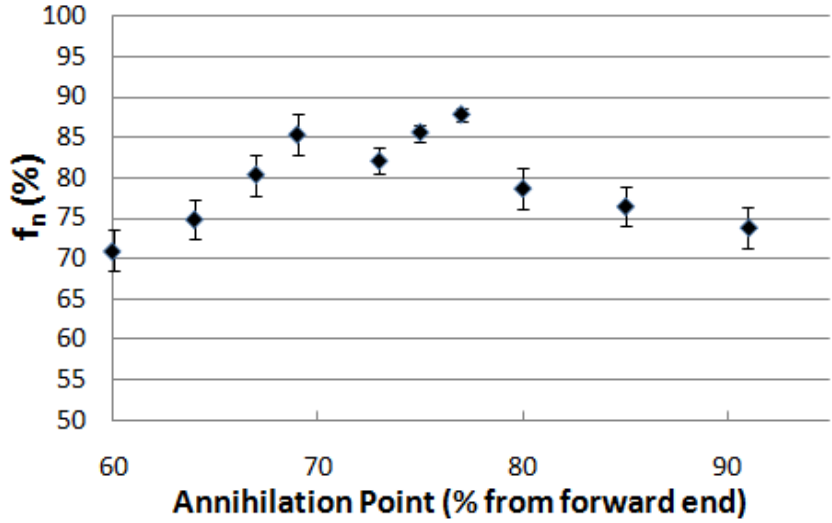

Fig 3: Nozzle efficiency as a function of $L_{annih}$, the position of the annihilation point along the axis of the nozzle. Here, $B_{min} = 0.325$ T and $B_{max} = 12.5$ T. The best performance corresponds to $L_{annih} \sim 77\%$.

Figs. 2 and 3 show result of scanning the remaining nozzle parameters. The most important observation is that the nozzle efficiency saturates at about 85% for $B_{max} \sim 12$ T. This is less than 10% of the maximum field assumed by Callas, although his coil was far smaller. Overall, the optimum magnetic field configuration indicated by the present study could be produced today. It is also important to note that Callas [12] published a conceptual sketch of a cone-shaped nozzle that had many similarities to the design reported in the present paper and would likely be capable of a comparable efficiency. The lower performance in the quantitative study of Callas was presumably a consequence of the limitations of the Monte Carlo simulation environment available in the late 1980s.

The event display feature provided by Geant is highly valuable in debugging and in independently verifying that charged particles follow the expected paths based on the field strength and direction in each region of space, and that the simulated nozzle is indeed achieving the efficiency (high or low) calculated by the code that reads and analyzes the Geant output. Fig. 4 shows a typical example with all nozzle parameters in the optimum region. From the trajectories observed over many events, it is clear that the dimensions $L_{coil} = 3.8$ m and $R_{coil} = 1.5$ m are sufficient, and that $L_{annih} = 77\%$ allows adequate space for particles emitted close to the thrust vector direction (to the left) to be turned around, while still ensuring that the field to the aft (right) of the annihilation point can bend charged particle trajectories relatively close to antiparallel to the thrust vector as they exit the nozzle. Finally, the event display reveals that particles forward of the annihilation point almost invariably remain well away from the coil radius, so a conical nozzle with smaller volume than the cylinder for the present study would have similar efficiency. However, a cylindrical nozzle geometry is less difficult to simulate.

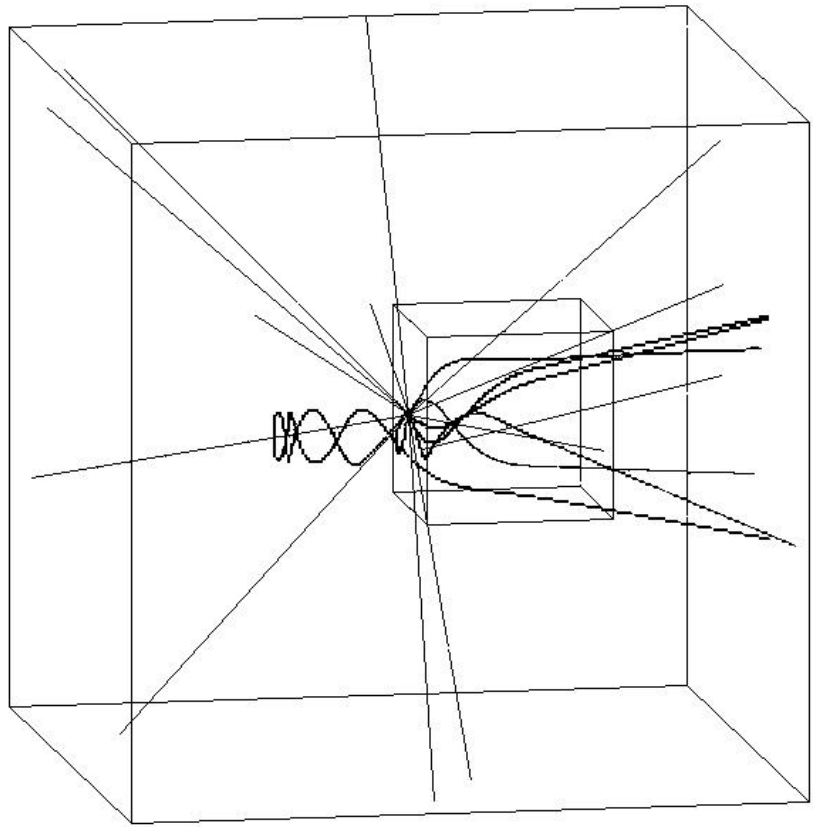

Fig 4: Event display from Geant4. For a better illustration of the different patterns of charged particle tracks, two separate annihilations are superimposed. The straight lines show uncharged particles (mostly gammas from neutral pion decay), while curved tracks are charged particles. The nozzle parameters here are the optimum ones already discussed. The inner fiducial cube has sides of 1.6 m, and the outer fiducial cube has sides of 6.4 m, with the annihilation point at its centre. These cubes are for orientation only, and their positions and dimensions have no significance for the simulation calculations. The forward end of the nozzle lies 2.9 m to the left of the annihilation point.

## 6. CONCLUSIONS FROM NOZZLE OPTIMIZATION

One overriding conclusion is that the Geant/CHIPS combination offers a viable and powerful simulation environment for the quantitative study of various design options for antimatter propulsion. In particular, a beamed core magnetic nozzle can deliver an efficiency $f_n$ more than double the best number from previous simulations. The most comprehensive overview of beamed core propulsion, by Frisbee [29-33], assumes an effective charged pion exhaust speed $v_e = \langle v_{\pi\pm} \rangle f_n = (0.92c)(0.36) = 0.33c$, whereas this study (using the updated $v_{\pi\pm}$ from Ref. [35]) indicates $v_e = (0.81c)(0.85) = 0.69c$.

## 7. CONNECTION TO MAXIMUM ATTAINABLE SPACECRAFT SPEED

The classic Tsiolkovsky formula can be written $\Delta v = v_e \ln (M_{initial}/M_{final})$, where $\Delta v$ is the change in speed imparted to a rocket, and $M_{initial}$ and $M_{final}$ are the mass with and without fuel, respectively. This expression must be generalized in the case of beamed core propulsion to allow for "loss of propellant" [29-33] – some of the fuel is expended on isotropic emission of uncharged particles, a crucial source of inefficiency that has no parallel in today's chemical rockets. For an engine powered by liquid hydrogen and liquid oxygen, typically $v_e = 4.4$ km/s, and vehicle speeds (before gravity assist) of more than $3.5v_e$ are attainable. With loss of propellant, maximum vehicle speeds are closer to $v_e$, even with multiple stages [32,33].

Frisbee's papers explain in depth the needed generalization to account for emission of uncharged particles [29-33]. When loss of propellant is taken into account, Frisbee has shown that $v_e \sim 0.3c$ leads to a beamed core rocket facing daunting challenges in reaching a true relativistic cruise speed on a one-way interstellar mission where deceleration at the destination (a "rendezvous" mission) would be involved. For example, Table 5 in Ref. [32] indicates that with a payload of 100 metric tons, a 4-stage beamed core rocket designed for a cruise speed of $0.42c$ on a 40 light-year rendezvous mission would require 40 million tons of antimatter fuel. If the cruise speed were limited to $0.25c$ or less, only two stages might be needed, and Frisbee envisaged viable interstellar missions with as few as one beamed core stage; in such scenarios, fuel requirements would be dramatically lower [32,33].

With the new reference point of $v_e = 0.69c$ provided by the present Geant-based simulation, true relativistic speeds once more become a possibility using the highest performance beamed core propulsion in the distant future. In this context, "true relativistic" refers to speeds $v$ where the Lorentz factor $\gamma = (1 - v^2/c^2)^{-1/2}$ exceeds unity by a significant amount.

An absorber/shield of neutral particles, located forward of the annihilation point, would have benefits for propulsion. Such a component would add to the forward thrust, although the absorbed neutral particles would of course be less effective in producing thrust than charged particles of the same momentum magnitude. Progress in this area would amount to an amelioration of the "loss of propellant" factor discussed above, and thus would give a modest boost to the attainable vehicle speeds. Although all calculations to date have ignored this aspect of optimizing beamed core propulsion, Geant is very well suited for a quantitative study of this topic.

## 8. LONG-TERM OUTLOOK FOR BEAMED CORE PROPULSION

The unmatched energy density of antimatter makes it an obvious fuel choice for the ultimate in advanced spacecraft propulsion. Antimatter has been a natural focus for the most futuristic and challenging missions, especially those venturing beyond the solar system. In any scenario where very limited availability of antimatter is not the overriding limitation, the highest performance would be achieved when the antimatter annihilation products generate thrust directly, after being deflected and collimated by an electromagnetic nozzle (the beamed core concept).

The prospect for spacecraft propulsion with antimatter as a fuel crucially depends on whether it ever becomes feasible to accumulate antimatter in macroscopic quantities and store it safely until needed. In 2009, Close published a book titled *Antimatter* [48], aimed at the general public. Close assumed that technology for antimatter production will remain static, and argued that it will take 1000 years to make a microgram of antimatter. In contrast, Frisbee made a prediction that the amount of bulk antihydrogen obtainable will grow exponentially, much like the growth of the intensity of beams of antiprotons at accelerators, which has increased about four orders of magnitude per decade since the discovery of the antiproton in the 1950s [32,33]. The same paper also draws attention to the growth in production of an important contemporary rocket fuel, namely liquid hydrogen, which has likewise followed an exponential pattern, but ironically with a longer time-constant than for the growth in intensity of antiproton beams. Based on the sparse data for neutral antihydrogen, Frisbee predicted that microgram quantities will be available by the middle of the 21st century [32,33]. Very recent research on trapping antihydrogen at CERN does indeed suggest a pattern of rapid progress [49-51].

Furthermore, production of antimatter for propulsion does not need to rely solely on the approach used today at accelerator labs, where proton-antiproton pairs are created in matter-on-matter collisions – this production method is extremely expensive and has very low energy efficiency, and is the main reason for skepticism by Close and others. An exciting new development was announced to the world in early August 2011 by the PAMELA collaboration [52]: the discovery of large fluxes of antiprotons trapped by the earth's magnetic field. Such trapping of antimatter by the earth and other planets had been predicted theoretically [53]. Following the installation of the Alpha Magnetic Spectrometer on the International Space Station in mid-2011, there will be an enhanced capability in the future to detect, identify and measure charged particles and antiparticles in earth orbit [54].

The recent PAMELA discovery, in which the observed antiproton flux is three orders of magnitude above the antiproton background from cosmic rays, paves the way for possible harvesting of antimatter in space. Theoretical studies suggest that the magnetosphere of much larger planets like Jupiter would be even better for this purpose [53]. If feasible, harvesting antimatter in space would completely bypass the obstacle of low energy efficiency when an accelerator is used to produce antimatter, and thus could offer a solution to the main difficulties stressed by the skeptics.

**ACKNOWLEDGEMENTS**

The authors acknowledge the European Center for Nuclear Research (CERN) for developing and maintaining the Geant software toolkit [36], and thank D. and J. Keane and D. M. Manley for discussion and assistance.